\documentclass[twocolumn,prl,superscriptaddress,showpacs,10pt,aps,draft]{revtex4-1}

\usepackage{color}

\usepackage{amsmath}
\usepackage{amssymb}
\usepackage{graphicx}
\usepackage{color}

\usepackage[T1]{fontenc}
\usepackage[utf8]{inputenc}

\usepackage{amsthm}
\usepackage{bbm}
\begin{document}
\title{Comment on ``Experimental Test of Error-Disturbance Uncertainty Relations by Weak Measurement''}

\author{Paul Busch}
\email{paul.busch@york.ac.uk}
\affiliation{Department of Mathematics, University of York, York, United Kingdom}

\author{Pekka Lahti}
\email{pekka.lahti@utu.fi}
\affiliation{Turku Centre for Quantum Physics, Department of Physics and Astronomy, University of Turku, FI-20014 Turku, Finland}

\author{Reinhard F. Werner}
\email{reinhard.werner@itp.uni-hannover.de}
\affiliation{Institut f\"ur Theoretische Physik, Leibniz Universit\"at, Hannover, Germany}

\date{\today}
%\begin{abstract}

%\end{abstract}
\maketitle
This paper \cite{Ozawa-etal2014} is the latest in a long series of theoretical and experimental works purporting to have demonstrated a violation of Heisenberg's so-called error-disturbance uncertainty relation. This claim, which originated with theoretical work of Ozawa around 2002, has stirred a considerable media hype since 2012, which would be justified if the claim were correct. But it is not, for the following reasons.

One has to recall at this point that Heisenberg's original discussion of the $\gamma$-ray microscope, in keeping with the aims he states in his introduction,  is intended only as a heuristic tool. He is intentionally vague, and states his relations not as inequalities but using a mathematically unexplained tilde. Although he promises a proof to be given later in the paper on the basis of his commutation relations he never gave one, nor ever stated his relation in a form sufficiently precise to even begin thinking of a proof. This is why finding precise quantum mechanical counterparts of Heisenberg's heuristic error and disturbance is not at all straightforward. But it is possible: We have given a natural definition of error and disturbance for which we proved an inequality of the usual form \cite{BLW2013c,BLW2013b}. This puts the error-disturbance tradeoff on the same level of quantitative rigour and generality as the usual uncertainty relation for the variances of position and momentum in the same state.

What Ozawa calls ``Heisenberg's error-disturbance relation'' -- inequality (1) in \cite{Ozawa-etal2014} -- is superficially of the same form, but he chooses different formal definitions of error and disturbance, which can be traced to the work of Arthurs and Kelly in the 60s \cite{AK65}. We have shown \cite{BLW2013a,BLW2014a} that these definitions have serious conceptual deficiencies undermining their interpretation as error and disturbance.
%Some of these problem can be traced to neglecting the consequences of quantum non-commutativity in an expression, which superficially looks like Gau\ss's definition of root mean square deviation between two random variables. %Since non-commutativity is closely related to uncertainty, we find this oversight particularly puzzling.
Moreover, it has been shown by simple counterexamples \cite{Appleby1998b}, long before Ozawa, that with these definitions a general uncertainty relation does not hold. This unspectacular observation has now repeatedly been verified experimentally with \cite{Ozawa-etal2014} the latest in the series. Similarly, groups in Toronto \cite{Roz12} and Brisbane \cite{Branciard2014} have reported experiments violating the  inequality that Ozawa wrongly attributes to Heisenberg. [In contrast to the Toronto group, the Brisbane researchers have {\em not} adopted this attribution.] But these experimental results do {\em not} help to refute Heisenberg's heuristics or the general idea of a quantitative error-disturbance tradeoff. They only show that the definitions for error and disturbance chosen by Ozawa, in addition to their intrinsic problems, 
are not suitable for the task of expressing such a tradeoff.

Ozawa has provided some additional terms, which turn his false inequality (1) into a correct one (Eq. (3) of [1]).  This inequality has recently been tightened in an interesting paper by Branciard \cite{Branciard2013}. Both of these inequalities have been confirmed in some of the experiments mentioned. In a recent letter \cite{Buscemi2014} Ozawa and his coauthors have also changed to a state-independent definitions more like ours. These are positive contributions to the research field of rigorous measurement uncertainty relations. We hope that in the future this field can concentrate on its scientific agenda and leave behind the misrepresentations by Ozawa's ``refuting Heisenberg'' campaign.

\bibliographystyle{unsrt}
%\bibliography{UR}

\end{document}